\documentclass[a4paper]{jpconf}
\usepackage{citesort}
\usepackage{amsmath}
\usepackage{amsfonts}
\usepackage{amssymb}
\usepackage{graphicx}

\def\ybco{YBa$_2$Cu$_3$O$_{7-x}$ }
\def\fe2o3{Fe$_2$O$_3$}

\begin{document}

\title{Effect of high-intensity ultrasound on superconducting properties of polycrystalline \ybco}

\author{T Prozorov}

\address{Materials Chemistry \& Biomolecular Materials, Ames Laboratory, Ames, IA 50011, U.S.A}

\ead{tprozoro@ameslab.gov}

\author{B McCarty and R Prozorov}
\address{Ames Laboratory and Department of Physics \& Astronomy, Iowa State University, Ames, IA 50011, U.S.A}

\ead{prozorov@ameslab.gov}

\begin{abstract}
High intensity ultrasonic irradiation (sonication) of alkane slurries of polycrystalline \ybco leads to a significant modification of the grain morphology and, if performed with enforced oxygen flow, results in the increase of the superconducting transition temperature. Sonication with added Fe(CO)$_5$ produces magnetic \fe2o3 nanoparticles deposited on the surface of \ybco (YBCO) granules. Upon sintering these nanoparticles should act as efficient pinning centers utilizing both condensation and magnetic contributions to the free energy. The developed method could become a major technique to produce practically useful high-pinning nanocomposite materials based on \ybco superconductor.
\newline\newline
\textit{Date: 14 June 2008}
\end{abstract}

The major challenge of applied superconductivity is to find a delicate balance between strong pinning and yet retain basic superconducting properties, such as transition temperature, T$_c$, and intergrain connectivity. Best candidates for applications are nanocomposite superconductors where pinning centers are introduced in a controlled way. There is a great deal of activity in this field. Tapes and films is a large and promising area of research \cite{tapes}. Others concentrate on improving bulk superconductors. In a typical scenario, additives (dopants, nanoparticles) are mixed into the pre-cursors or simply into powdered initial material. The composite is then sintered at high temperature and/or pressure to produce the final strong-pinning superconductor \cite{hightc,bulk}. Our research shows that high-intensity ultrasound can be efficiently used to produce nanocomposite superconducting materials \cite{snezhko2005}. Enhanced pinning properties of MgB$_2$ \cite{Prozorov2003a} and Bi$_2$Sr$_2$CaCu$_2$O$_{8+x}$ \cite{Prozorov2004c} superconductors have been reported. Recently, high - temperature isostatic pressure treatment of sonicated precursors resulting in a dense nanocomposite MgB$_2$ superconductors has been demonstrated \cite{brett2008}.

In this contribution we show that sonochemically treated \ybco (YBCO) is a promising main component for such superconducting nanocomposite. Successful deposition of \fe2o3 nanoparticles on the surfaces of individual YBCO granules is demonstrated. Importantly, the process does not result in the deterioration of the basic superconducting properties. Moreover, when sonication was performed under partial oxygen flow, the superconducting transition temperature of the resulting material had increased.

\begin{figure}[t]
\includegraphics[width=7cm]{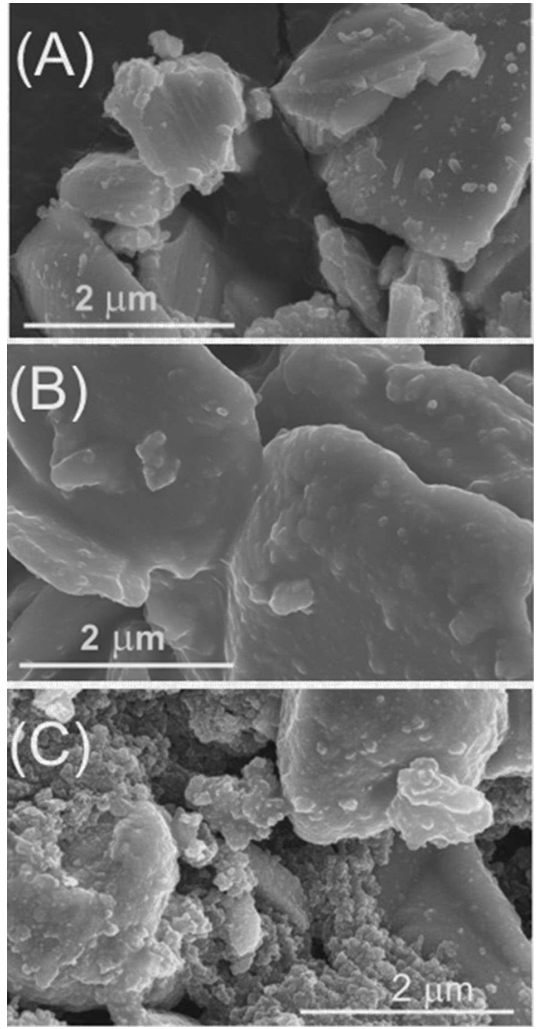}\hspace{1.5pc}%
\begin{minipage}[b]{8.2cm}\caption{\label{fig1}\ybco 1-6 $\mu$m powder from Alfa Aesar) was used as received. Pentane and decane (anhydrous, 99+\%, Aldrich) were distilled under argon prior to use. Polycrystalline \ybco was ultrasonically irradiated for 120 min at 263 K in 20 mL of decane in an open flask under moderate argon flow (20 mL/min), using direct immersion ultrasonic horn (Sonics VCX-750 at 20 kHz and 50 W/cm$^2$). Various slurry loadings were used.  A similar set of slurries was sonicated under ambient atmosphere with the addition of different amounts of Fe(CO)$_5$. Oxygen content in the starting material was probed by the iodometric titration. To maintain the necessary oxygen content during the sonochemical irradiation of YBCO powder, and to explore effects of sonochemical conditions on T$_c$, sonication was performed in 2\% (w/w) slurry in ethylene glycol, under 40:20 Ar:O$_2$ flow.  All ultrasonically treated materials were collected by filtration, washed with dry pentane, and air-dried overnight. The resulting dry powders were pelletized at room temperature at a pressure of 2 GPa for 24 hours.
The figure shows:  (A) Commercial powder. (B) YBCO sonicated in 2\% (w/w) decane slurry at 263 K. (C) \fe2o3 nanoparticles, distributed over the grains of irradiated YBCO  after sonication with 180 $\mu$mol of Fe(CO)$_5$. }
\end{minipage}
\end{figure}

Powder morphology was studied by using Hitachi S-4700 and FEI Quanta 200 SEMs. Particle size was analyzed on a Phillips CM12 TEM equipped with EDX. Surface chemical composition of the modified powders was monitored by using XPS and microprobe EDX. XPS analysis was conducted on a Physical Electronics PHI 5400 X-Ray Photoelectron Spectrometer. Powder X-ray Diffraction study was performed by using a Rigaku D/MAX diffractometer. Magnetic measurements were conducted using a Quantum Design MPMS-5 magnetometer.

\begin{figure}[t]
\includegraphics[width=10cm]{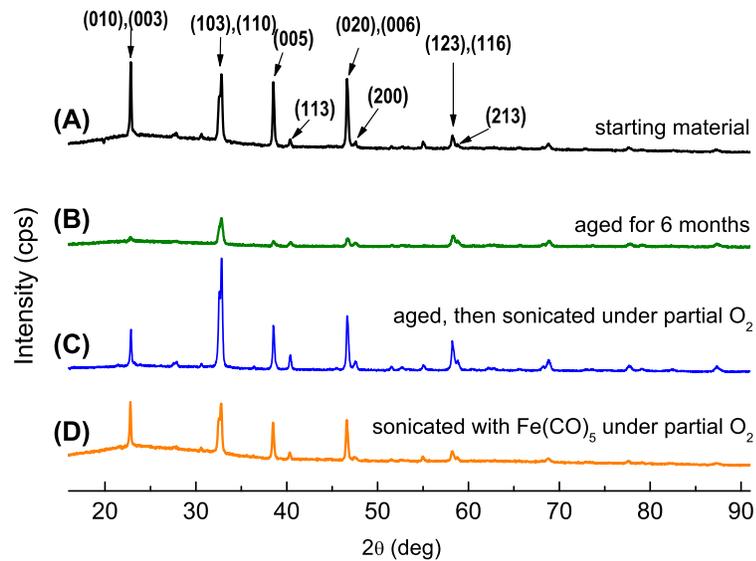}\hspace{1.5pc}%
\begin{minipage}[b]{5cm}\caption{\label{fig2}Powder x-ray diffractograms (XRD) of polycrystalline YBCO. (A) Starting (as received) material. (B) XRD pattern of YBCO aged for 6 months under ambient conditions. (C) Aged YBCO sonicated under partial O$_2$ flow. (D) YBCO powder co-sonicated with Fe(CO)$_5$.}
\end{minipage}
\end{figure}

SEM images of sonochemically treated YBCO powders shown in Fig.~\ref{fig1} (B)(C) reveal substantial fusion, smoothing, and improved interconnecting of individual grains, as compared to the starting material, Fig.~\ref{fig1}(C). Figure \ref{fig2} shows powder XRD spectra of several powders used in this study. Noticeably, material that spent 6 months on the shelf shows significant degradation. However, when sonicated with partial O$_2$ flow, it was not only fully restored, but became better indicating that such treatment is similar to the annealing in oxygen atmosphere.

\begin{figure}[b]
\begin{center}
\includegraphics[width=9cm]{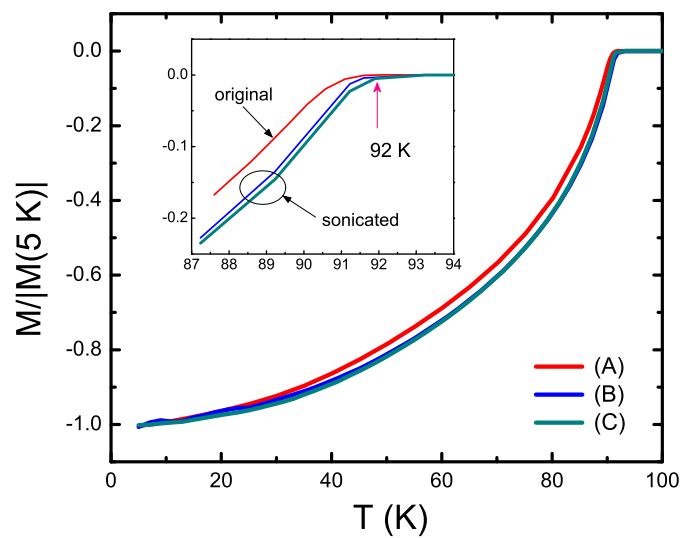}
\end{center}
\caption{\label{fig3}Effect of oxygen flow during sonication on the $M(T)$ curves measured in resulting YBCO powder. (A) starting material. (B) YBCO  sonicated under partial oxygen pressure. (C) YBCO sonicated under partial oxygen pressure with 9 mmol Fe(CO)$_5$. Sonication of 2 \% (w/w) ethylene glycol slurry at 263 K, 20 kHz, 50 W/cm$^2$. Inset:  vicinity of T$_c$.}
\end{figure}

To further explore this observation and examine the effect on superconducting properties, figure \ref{fig3} shows the M vs. T curves of sonicated YBCO. Compared to co-sonicated with O$_2$, as received powder shows smaller T$_c$ and superconducting screening.

\begin{figure}[t]
\includegraphics[width=8.5cm]{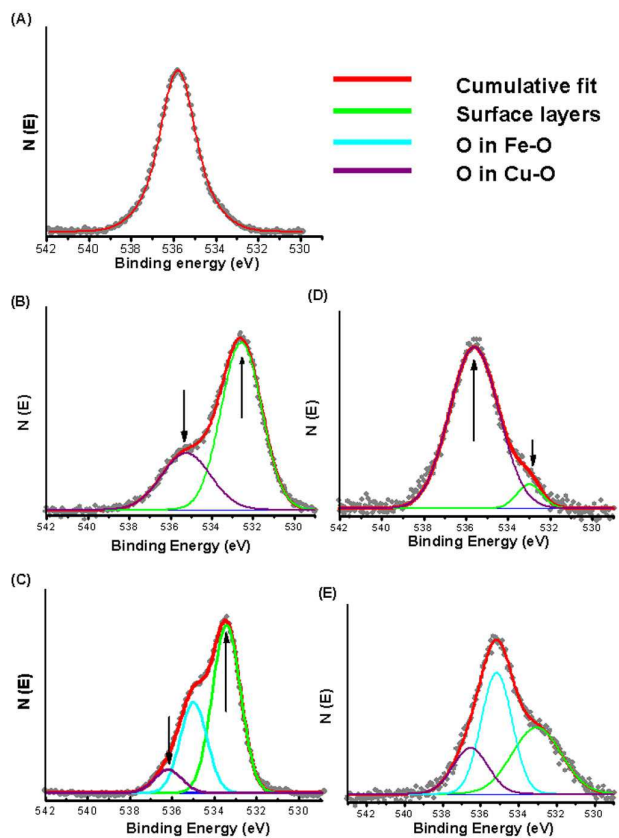}\hspace{2pc}%
\begin{minipage}[b]{6.5cm}\caption{\label{fig4}Comparison of XPS peaks on O$^{1s}$ site in several YBCO samples. (A) Starting material shows relatively narrow single oxygen peak. (B) Low-energy O$^{1s}$ peak appears in the sample sonicated under a partial oxygen flow. The position of the peak matched that of samples sonicated under the Ar. (C) YBCO sonicated under air/Ar flow with 4.5  mmol Fe(CO)$_5$. (D) YBCO sonicated under O$_2$/Ar flow; (E) aged YBCO sonicated in ethylene glycol under O$_2$/Ar flow with 4.5 mmol of Fe(CO)$_5$. Evidently, sonication under partial oxygen flow with small amount of Fe(CO)$_5$ leads to disruption of oxygen surface layer, as indicated by the appearance of the intermediate  peak. These observations indicate that sonication in oxygen allowed control over the surface oxygen concentration.}
\end{minipage}
\end{figure}

In Fig.~\ref{fig4}, XPS O$^{1s}$ spectra for several \ybco  samples are compared.  The starting material

In conclusion, irradiation with high-intensity ultrasound of alkane slurries containing granular superconductors results in a substantial change of powder's morphology without affecting its bulk chemical composition. In particular, sonochemical treatment of polycrystalline \ybco high-T$c$ superconductor showed a dramatic effect on morphology and superconducting properties. Decane slurries of polycrystalline \ybco were ultrasonically irradiated with different slurry loadings and duration of the sonochemical treatment. To maintain the oxygen content in sonochemically treated \ybco, sonochemical irradiation of slurries was performed under a partial oxygen flow in decane and in ethylene glycol. The sonication resulted in significant modification of grain morphology and improved intergrain coupling. Effectiveness of sonication increases with the decrease of slurry loading, due to more effective interparticle collisions.  Novel nanocomposites with finely dispersed ferrimagnetic nanoparticles have been successfully prepared. Obtained materials show enhanced Meissner screening and larger magnetic irreversibility. Critical temperature T$_c$ is improved due to oxygen restoration in the grain outer layers during sonication in a partial O$_2$ flow that creates local oxygen plasma. The sonochemical method and post-sonochemical annealing are to be further optimized to achieve maximal pinning enhancement in granular \ybco.

\ack
Work at the Ames Laboratory was supported by the Department of Energy-Basic Energy Sciences under Contract No. DE-AC02-07CH11358. R. P. acknowledges support from NSF grant number DMR-06-03841 and the Alfred P. Sloan Foundation.

\end{document}